
%
%
%
%
\input harvmac
%
%
%
\def\listrefs{\footatend\bigskip\bigskip\immediate\closeout\rfile\writestoppt
\baselineskip=12pt\parskip=2pt plus 1pt
\noindent{{\bf References}}\medskip{\frenchspacing%
\parindent=20pt\escapechar=` \input \jobname.refs\vfill\eject}\nonfrenchspacing}
%
%
%
\def\RF#1#2{\if*#1\ref#1{#2.}\else#1\fi}
\def\NRF#1#2{\if*#1\nref#1{#2.}\fi}
\def\refdef#1#2#3{\def#1{*}\def#2{#3}}
%
%
%
\def\ts{\hskip .16667em\relax}

\def\IJMP{{\it Int.\ts J.\ts Mod.\ts Phys.\ts}}
\def\JMP{{\it J.\ts Math.\ts Phys.\ts}}
\def\JP{{\it J.\ts Phys.\ts}}
\def\NP{{\it Nucl.\ts Phys.\ts}}
\def\PL{{\it Phys.\ts Lett.\ts}}

\def\PRL{{\it Phys.\ts Rev.\ts Lett.\ts}}

\def\SJNP{{\it Sov.\ts J.\ts Nucl.\ts Phys.\ts}}
\def\Zm{Zamolodchikov}
\def\AZm{A.\ts B.\ts \Zm}
\def\AlZm{Al.\ts B.\ts \Zm}
\def\me{P.\ts E.\ts Dorey}
\def\dur{H.\ts W.\ts Braden, E.\ts Corrigan, \me\ and R.\ts Sasaki}
%
%
\refdef\rABLa\ABLa{C.\ts Ahn, D.\ts Bernard and A.\ts LeClair, \NP
{\bf B346} (1990) 409}

\refdef\rAFZa\AFZa{A.\ts E.\ts Arinshtein, V.\ts A.\ts Fateev and
 \AZm, \PL {\bf B87} (1979) 389}

\refdef\rBCa\BCa{R. Brunskill and A. Clifton-Taylor, {\it English Brickwork}
 (Hyperion 1977)}

\refdef\rBCDSf\BCDSf{\dur, \NP {\bf B338} (1990) 689;
 \NP {\bf B356} (1991) 469}

\refdef\rBRa\BRa{V.\ts V.\ts Bazhanov and N.\ts Reshetikhin, \JP {\bf
A23} (1990) 1477}

\refdef\rCMe\CMe{P.\ts Christe and G.\ts Mussardo, \NP {\bf B330} (1990)
465; \IJMP {\bf A5} (1990) 4581}

\refdef\rDf\Df{\me, \NP {\bf B358} (1991) 654; \NP {\bf B374} (1992) 741}

\refdef\rDDa\DDa{C.\ts Destri and H.\ts J.\ts de Vega, {\it Phys. Lett.}
 {\bf B233} (1989) 336}

\refdef\rDDb\DDb{C.\ts Destri and H.\ts J.\ts de Vega, \NP {\bf B358}
(1991) 251}

\refdef\rDGPZa\DGPZa{G.\ts Delius, M.\ts Grisaru, S.\ts Penati and D.\ts
Zanon, \PL {\bf B256} (1991) 164\semi \NP {\bf B359} (1991) 125}

\refdef\rDGZa\DGZa{G.\ts W.\ts Delius, M.\ts T.\ts Grisaru and D.\ts Zanon,
preprints CERN-TH.6333/91 and CERN-TH.6337/91}

\refdef\rDZa\DZa{P.\ts Di\ts Francesco and J.-B.\ts Zuber, \NP {\bf B338}
(1990) 602}

\refdef\rEYa\EYa{T. Eguchi and S-K Yang, {\it Phys. Lett.} {\bf  B224} (1989)
 373}

\refdef\rFb\Fb{M.\ts D.\ts Freeman, \PL {\bf B261} (1991) 57}

\refdef\rFc\Fc{P.\ts Fendley, `Excited-state thermodynamics', preprint
BUHEP-91-16}

\refdef\rFLTa\FLTa{Lectures by L.D.\ts Faddeev, J.H.\ts Lowenstein and
H.B.\ts Thacker, in {\it Advances in Field Theory and Statistical
Mechanics}, Proceedings of the 1982 Les Houches Summer School --
J.-B.\ts Zuber and R.\ts Stora Eds}

\refdef\rFZa\FZa{V.\ts A.\ts Fateev and \AZm, {\it Int. J. Mod. Phys.}
 {\bf A5} (1990) 1025}

\refdef\rFZb\FZb{V.\ts A.\ts Fateev and \AlZm,
 \PL {\bf B271} (1991) 91}

\refdef\rFZc\FZc{V.\ts A.\ts Fateev and \AZm, \NP {\bf B280} 644}

\refdef\rGc\Gc{D.\ts Gepner, \NP {\bf B290} (1987) 10}

\refdef\rHd\Hd{T.\ts J.\ts Hollowood, `Quantizing $SL(N)$ solitons and
the Hecke algebra', Oxford preprint OUTP-92-03P}

\refdef\rHMa\HMa{T.\ts J.\ts Hollowood and P.\ts Mansfield, \PL {\bf B226}
 (1989) 73}

\refdef\rKa\Ka{M.\ts Karowski, \NP {\bf B153} (1979) 244}

\refdef\rKg\Kg{A.\ts N.\ts Kirillov, {\it Zapiski Nauch. Semin. LOMI}
{\bf 164} (1987) 121}

\refdef\rKMa\KMa{T.\ts R.\ts Klassen and E.\ts Melzer, \NP {\bf B338}
(1990) 485}

\refdef\rKMb\KMb{T.\ts R.\ts Klassen and E.\ts Melzer, \NP {\bf B350}
(1991) 635}

\refdef\rLc\Lc{L.\ts Lewin, Dilogarithms and associated functions
(Macdonald 1958)}

\refdef\rMe\Me{M.\ts J.\ts Martins, \PL {\bf B240} (1990) 404}

\refdef\rMh\Mh{M.\ts J.\ts Martins, \PRL {\bf 67} (1991) 419}

\refdef\rMk\Mk{M.\ts J.\ts Martins, \PL {\bf B277} (1992) 301}

\refdef\rMl\Ml{M.\ts J.\ts Martins, SISSA preprints SISSA-EP-72,
SISSA-EP-85}

\refdef\rMn\Mn{P.\ts Mansfield, \PL {\bf B242} (1990) 387}

\refdef\rRb\Rb{F.\ts Ravanini, \PL {\bf B282} (1992) 73}

\refdef\rYYa\YYa{C.\ts N.\ts Yang and C.\ts P.\ts Yang, \JMP {\bf 10}
(1969) 1115}

\refdef\rZa\Za{\AZm, ``Integrable Field Theory from Conformal Field Theory",
 Proceedings of the Taniguchi Symposium, Kyoto (1988)}

\refdef\rZb\Zb{\AZm, {\it Int. J. Mod. Phys.} {\bf A4} (1989) 4235}

\refdef\rZc\Zc{\AZm, {\it JETP Letters} {\bf 43} (1986) 730}

\refdef\rZd\Zd{\AZm, {\it Int. J. Mod. Phys.} {\bf A3} (1988) 743}

\refdef\rZe\Ze{\AZm, {\it Sov. Sci. Rev., Physics}, {\bf v.2} (1980)}

\refdef\rZf\Zf{\AZm, {\it Teor. Mat. Fiz.} {\bf 65} (1985) 347}

\refdef\rZg\Zg{\AlZm, \NP {\bf B242} (1990) 695}

\refdef\rZh\Zh{\AZm, Kiev IMP preprint 87-65P (1987)}

\refdef\rZi\Zi{\AlZm, \PL {\bf B253} (1991) 391}

\refdef\rZj\Zj{\AlZm, \NP {\bf B348} (1991) 619}

\refdef\rZk\Zk{\AlZm, \NP {\bf B358} (1991) 497}

\refdef\rZl\Zl{\AlZm, \NP {\bf B358} (1991) 524}

\refdef\rZn\Zn{\AlZm, \NP {\bf B366} (1991) 122}

\refdef\rZo\Zo{\AlZm,
`Resonance factorized scattering and roaming trajectories', preprint
ENS-LPS-335, 1991}

\refdef\rZp\Zp{\AZm, \SJNP {\bf 44} (1986) 529}

\refdef\rZt\Zt{\AZm, Talk given in Oxford, January 1989}

\refdef\rZz\Zz{\Za\semi\Zb}

\refdef\rZZa\ZZa{\AZm\  and \AlZm, {\it Ann. Phys.}
 {\bf 120} (1979) 253}

\refdef\rZZb\ZZb{\AZm\  and \AlZm, `Massless factorized scattering and
sigma-models with topological terms', preprint ENS-LPS-355, 1991}

\def\bar{\overline}
\def\hat{\widehat}
\def\tilde{\widetilde}
\def\({\left(}
\def\){\right)}
\def\[{\left[}
\def\]{\right]}

\def\th{^{\rm th}}

\def\exp{{\rm exp}}
\def\min{{\rm min}}

\def\CC{{\cal C}}
\def\CI{{\cal I}}
\def\tCI{\tilde\CI}
\def\CJ{{\cal J}}
\def\tCJ{\tilde\CJ}

\def\CT{{\cal T}}
\def\CR{{\cal R}}

\def\ubl#1{\{#1\}}
\def\usbl#1{\(#1\)}

\def\Z{{\bf Z}}

\def\a{\alpha} \def\b{\beta} \def\g{\gamma} 
\def\ep{\varepsilon}  \def\l{\lambda} \def\p{\phi}
\def\t{\theta}   
\def\G{\Gamma}
  
\def\hm{\hat m}

\def\dilog{{\rm Li}_2}
\def\rdilog{{\cal L}}
\def\strutdepth{\dp\strutbox} 
\def\probsymbol{\vtop to \strutdepth{\baselineskip\strutdepth
  \vss\llap{\bf ??~~}\null}}
\def\prob#1{\strut\vadjust{\kern-\strutdepth\probsymbol}{\bf #1}}
\def\coset#1#2{$g^{(#1)}{\times}g^{(1)}/g^{(#2)}$}
\def\ccoset#1#2#3{$g^{(#1)}{\times}g^{(#2)}/g^{(#3)}$}
\def\intt{\int_{-\infty}^{\infty}}
\def\ri{{\rm I}}
\def\rii{{\rm II}}
\def\riii{{\rm III}}
\def\riv{{\rm IV}}
\def\rv{{\rm V}}
\def\tt{\thinspace}
\def\forget#1{}
\Title{\vbox{\baselineskip12pt\hbox{SPhT/92-065}\hbox{DFUB-92-09}}}%
{Staircase Models from Affine Toda Field Theory}
\centerline{
Patrick
E.~Dorey$^\dagger$\footnote{}{$^\dagger$~~dorey@poseidon.saclay.cea.fr}}
\medskip\centerline{Service de Physique Th\'eorique de
Saclay$^*$,\footnote{}{$^\ddagger$~~ravanini@bologna.infn.it}}
\centerline{91191 Gif-sur-Yvette cedex, France}
\bigskip
\centerline{
Francesco Ravanini$^\ddagger$
\footnote{}{$^*$~~{\it Laboratoire de la Direction des Sciences
de la Mati\`ere du Commissariat \`a l'Energie Atomique}}}
\medskip\centerline{INFN -- Sez. di Bologna,}
\centerline{Via Irnerio 46,}
\centerline{I-40126 Bologna, Italy}
\vskip .5in
\baselineskip=15pt plus 2pt minus 1pt
We propose a class of purely elastic scattering theories generalising the
staircase model of Al.\ts B.\ts Zamolodchikov, based on the affine Toda
field theories for simply-laced Lie algebras $g=A,D,E$ at suitable
complex values of their coupling constants. Considering their Thermodynamic
Bethe Ansatz equations, we give analytic arguments in support of a
conjectured renormalisation group flow visiting the neighbourhood of each
$W_g$ minimal model in turn.
\Date{June 1992}

\newsec{Introduction}

The proposals of A.\ts B.\ts Zamolodchikov for the scaling region of
the Ising model in a magnetic field\ts\RF\rZz\Zz\ have stimulated a renewal
of interest in two-dimensional integrable field theory. To think of a
model as a perturbation of a conformal field theory
by certain of its relevant operators allows much non-perturbative
information to be extracted, starting with the existence of the
conserved charges which guarantee integrability, and often leading on from
there to hypotheses for an exact S-matrix and mass spectrum,
hypotheses which in many cases have now withstood
independent checks.

One such check is to `complete the circle', reconstructing information
on the ultraviolet limit from
the infra-red data contained in the hypothesised exact S-matrix. If
this matches with the content of the initial conformal field theory,
a very non-trivial confirmation of the whole
approach has been achieved. Better still, non-perturbative information
on the whole renormalisation group trajectory between ultraviolet and
infra-red can often be extracted. The machinery employed is known as
the Thermodynamic Bethe Ansatz (TBA)\ts\RF\rYYa\YYa, first used in this
(relativistic) context by Al.\ts B.\ts Zamolodchikov\ts\RF\rZg\Zg;
subsequently these techniques have been developed by many others.

The method is straightforward if the scattering is purely
elastic\ts\NRF\rMe\Me\NRF\rKMa\KMa\NRF\rKMb\KMb\refs{\rZg{--}\rKMb},
that is if the S-matrix is diagonal in a
suitable basis. The thermodynamics of the system in a box
can be reconstructed
from a set of coupled non-linear integral equations (the TBA
equations), and these equations
follow immediately from the logarithmic derivative of the S-matrix
itself. If the box is small enough, UV effects dominate and (to
mention one example) the central charge of the original conformal
field theory can be recovered.
Interesting features have been observed in the study of
some of these systems of equations, pointing to the possibility of further
exact results\ts\RF\rZi\Zi: the TBA equations for
models described by the so-called minimal ADE affine Toda S-matrices
(see below) can be written in a universal way, depending only on
corresponding non-affine Dynkin diagrams.

Unfortunately, the majority of cases involve non-diagonal S-matrices.
To derive TBA equations directly from such scattering data, the
so-called `colour transfer matrix' should be diagonalised\ts\RF\rZk\Zk, a
forbidding problem -- requiring the higher-level Bethe ansatz (see,
for example, ref.\ts\RF\rFLTa\FLTa\ts )
-- in many cases. Furthermore, many perturbations of conformal field
theories lead to flows, not to massive scattering in the
IR limit, but to a non-trivial fixed point so that the RG flow
interpolates between two conformal field theories. The
$\phi_{13}$ flows between minimal models\ts\RF\rZc\Zc\
provide the first examples
of this phenomenon. Despite some suggestive
proposals\ts\NRF\rZl\Zl\NRF\rZZb\ZZb\refs{\rZl,\rZZb}\
the S-matrix approach is much
more delicate in such cases, owing to the well-known difficulties of
treating massless particles in 1+1 dimensions.

An alternative strategy, developed by Al.\ts B.\ts Zamolodchikov in a
series of papers\ts\NRF\rZn\Zn\refs{\rZk,\rZl,\rZn},
is to start with an educated guess for the
relevant system of TBA equations, and then to check that they
correctly reproduce a reasonable number of parameters typical of the
UV theory, such as the central charge and the dimension of the perturbing
operator. Once it has been established to general satisfaction that
the proposed TBA is indeed associated with the perturbed conformal
field theory under discussion,
the RG flow of the vacuum energy (and possibly also of the energies of some
of the excited states\ts\RF\rMh{\Mh\semi\Fc} )
can be followed numerically, or IR data can be reconstructed
analytically, allowing the IR limit to be identified -- perhaps
with some other conformal field theory.

In this way Al.\ts B.\ts Zamolodchikov was able to propose TBA
systems for the flows of minimal models perturbed by
$\phi_{13}$, the least relevant operator.
Rather remarkably, these TBA systems can also be encoded on graphs
(in fact the $A_n$ Dynkin diagrams). In the diagonal TBA, energy terms
in the equations, one for each particle type, were naturally
attached to the nodes of the corresponding Dynkin diagram, as will
be explained further below. Here, the
structure is a little different, with all but one or two of the energy
terms being  set to zero. The correct distribution to take
depends on the sign of the perturbing parameter, and has a dramatic
effect on the predicted IR behaviour: for one choice the flow is
to a massive theory of kinks\ts\rZk, while the other gives the
flow to the next-lowest minimal model\ts\rZl.
Subsequent work\ts\rZn, further modifying
the distribution of energy terms, gave ans\"atze (with
identical diagrammatic structure) for TBA systems describing the flows
between $su(2)^k{\times}su(2)^l/su(2)^{k+l}$ coset
models.

Proceeding further, it was natural to look for TBA systems to describe
analogous perturbations of \ccoset k l {k+l}\ coset models, for which
$\phi_{13}$ becomes $\phi_{(id,id,adj)}\,$.
These were
given for $g=A_n$, $l=1$ in \RF\rMk\Mk, and then for the general
(simply-laced) case in \RF\rRb\Rb.
In this latter reference a universal form of the general TBA equations
was given, allowing the generalisation of the diagrammatic structure
found in \rZi\ to be identified: the relevant graph is a
`product' of the Dynkin diagram of $g$ (which had already
been seen to be relevant for the diagonal Toda-type S-matrices) with an
$A_k$ Dynkin diagram. Otherwise stated, the graph is formed by taking
a tower of $k$ copies of the $g$ Dynkin diagram, with links in the
vertical direction joining the replicated nodes. Presumably this
replication encodes the colour structure of the non-diagonal
S-matrices for these models\ts\RF\rABLa\ABLa, although the details remain
to be worked out.

For the minimal models ($g=su(2)$), there is another intriguing
result. Al.\ts B.\ts Zamolodchikov recently proposed\ts\RF\rZo\Zo\
a simple diagonal scattering theory (the so-called `staircase model'),
from which a TBA system followed in the standard
way. However, numerical study of the solutions of this system revealed
a structure much richer than any previously observed, with hoppings of
the vacuum energy (which is proportional to the effective central
charge) strongly suggestive of an RG flow for an underlying field
theory that passes by a sequence of $c<1$ (minimal model) fixed points.

In fact, Zamolodchikov's S-matrix is an analytic continuation of that
for the $A_1$ affine Toda field theory. This suggests a natural
generalisation to the other affine Toda field theories, and our aim in
this paper is to present arguments supporting the claim
that such S-matrices yield new staircase models which hop between
the \coset k {k+1}\ ($W_g$) coset conformal theories. The above-mentioned
structure of a tower of Dynkin diagrams for the TBA systems
found in \rRb\ emerges as a natural consequence of the block structure
of the affine Toda S-matrices.

While we were completing the writing of this paper, independent
work by Martins appeared\ts\RF\rMl\Ml, proposing the same generalisation
of \rZo\ as we give here, and backing this up with a detailed numerical
study of the solutions of the TBA system.
This seems to us to complement our rather more analytic approach,
and we therefore refer the reader to \rMl\ for
numerical confirmations of the various approximations and hypotheses
to be made below.

The paper is organised as follows: in section 2 we recall the
necessary features of the affine Toda S-matrices, and their
relationship with the root systems of the simply-laced Lie algebras.
The connection between the `minimal' and `full' affine Toda
S-matrices is described, and this leads to a key relationship between
their logarithmic derivatives. Section 3 presents the relevant
TBA equations, and discusses the `locality' structure of these
equations as a function of the coupling constant. For a suitable
analytic continuation of the coupling constant, the system becomes
highly non-local and involves two scales (rather than the more usual
one), the interaction of which is responsible for the intricate
hopping structure of the solutions. Effective central charges are predicted
for various asymptotic limits of these two scales, providing evidence that
the RG behaviour being described is indeed a roaming trajectory
visiting a sequence of $W_g$-minimal models. At the risk of
repeating material that may be well-known to experts in the field, we
have decided to make this part of the discussion reasonably explicit.
Finally, we give a few speculations as to the physical meaning of the
structures observed.

\newsec{The affine Toda S-matrices}

For real values of the coupling constant, the simply-laced affine Toda
theories seem to have a number of particularly simple
properties\ts\NRF\rAFZa\AFZa\NRF\rCMe\CMe\NRF\rBCDSf\BCDSf%
\NRF\rDDa\DDa\refs{\rAFZa{--}\rDDa}.
There are $r$ different types of particles, $r$ being the
rank of the associated Lie algebra. Particles
of different types are distinguished one
from another by conserved charges of non-zero spin, constraining
the S-matrix to be diagonal --- the scattering theory is
purely elastic.
Furthermore (and in contrast to the situation for non simply-laced
algebras) the mass ratios remain fixed at
their classical values, and do not pick up any coupling-constant
dependence via quantum effects. As a result, the full S-matrix elements
factorise into a product of two pieces: a minimal factor,
independent of the coupling constant, which contains the
physical-strip poles and thus all information on mass ratios and
bound-state structure; and a second `Z-factor' incorporating the
coupling-constant dependence. This piece simply introduces some
zeroes into the physical strip, ensuring that as the coupling
constant tends to zero the S-matrix tends to the identity, but
changes neither the bound-state structure nor the resulting bootstrap
equations. The situation for non simply-laced theories is markedly
more complicated\ts\RF\rDGZa\DGZa, and will not be discussed below.

Despite their apparent independence, both factors must satisfy the
same set of bootstrap equations, equations which follow from the pole
structure of the minimal piece. As a result, their functional forms are
very similar, a fact which will be relevant below. This is most clearly
seen when the S-matrix elements are written as a product of the
functional building blocks $\ubl{x}_B$ employed in ref.\ts\rBCDSf. First,
define elementary blocks $\usbl{x}$ by
\eqn\usbldef{
\usbl{x}(\t)={\sinh\bigl({\theta\over 2}+{i\pi x\over 2h}\bigr)\over
        \sinh\bigl({\theta\over 2}-{i\pi x\over 2h}\bigr)}~,}
where $h$ is the Coxeter number of the algebra involved. (Other
authors use the notation $f_{\a}(\t)$ for $\usbl{h\a}(\t)$; including
the factor of $h$ in the notation, as here, has the advantage that
the parameter $x$ then turns out to take integer values.)
The bigger block, incorporating the coupling-constant dependence via
the function $B(\b)=2\b^2/(\b^2{+}4\pi)$, is
\eqn\ubldef{\ubl{x}_B={\usbl{x-1}\usbl{x+1}\over\usbl{x-1+B}\usbl{x+1-B}}~.}
The factorisation of this block into minimal and coupling-constant
dependent pieces is clear, as is the relationship between the two
factors. To express this precisely, define a symmetrised shift
operator\ts\rCMe, acting on functions $f(\t)$, by
\eqn\Rdef{(\CR_yf)(\t)=f(\t{-}i\pi y/h)f(\t{+}i\pi y/h).}
Then
\eqn\Rprop{\CR_y\usbl{x}=\usbl{x-y}\usbl{x+y},}
and so
\eqn\ubldeft{\ubl{x}_B={\CR_1\usbl{x}\over\CR_{1-B}\usbl{x}}.}
Thus the only difference between the two factors in
the building block is a reciprocation and a change in the
shift. This property is inherited by
the full S-matrix elements, though it can sometimes be hidden due to the
relations $\usbl{0}= -\usbl{h} =1$.

{}From the point of view of the bootstrap equations, it is perfectly
consistent to consider the minimal part on its own, as removing the
Z-factor leaves the physical poles and the signs of their residues
unchanged. Such S-matrices are products of the corresponding
minimal blocks,
\eqn\mubldef{\ubl{x}_{\min}=\usbl{x-1}\usbl{x+1}=\CR_1\usbl{x}~.}
However, the physical interpretation is very different: TBA
calculations\ts\refs{\rZg{--}\rKMa}
have established beyond reasonable doubt that
these S-matrices correspond to perturbations of
\coset 1 2\ coset conformal field theories.

Although the details are not essential for the main development, it is
worth recalling how these S-matrix elements can be written in a
uniform way for all the ADE series\ts\RF\rDf\Df. The construction uses
properties of a Coxeter element $w$ taken from the Weyl group of the
relevant root system $\Phi$, each particle species $a$ being
associated with an orbit $\G_a\subset\Phi$ of $w$. Letting
$\G_a^+=\G_a\cap\Phi^+$ denote the intersection of such an orbit with
the positive roots,
\eqna\myS
$$\eqalignno{ S^B_{ab}&= \prod_{\b\in\G_b^+}
  \ubl{u(\phi_a,\beta)+1}_B^{(\l_a,\b)}\,;&\myS a\cr
 S^{\min}_{ab}&= \prod_{\b\in\G_b^+}
  \ubl{u(\phi_a,\beta)+1}_{\min}^{(\l_a,\b)}\,.&\myS b\cr}$$
Here, $\l_a$ is the fundamental weight associated with the $a\th$ spot
on the Dynkin diagram, $\phi_a{=}(1{-}w^{-1})\l_a$ is a particular root
in the orbit $\G_a$, and $2\pi u(\phi_a,\b)/h$ is the angle between
the projections of $\phi_a$ and $\b$ into the $e^{2\pi
i/h}$-eigenplane of $w$. For more explanation see \rDf, but really it
is only the general form of \myS{}\ that will be relevant to what
follows.

At the level of the S-matrices, the factorisation seen for the blocks
in \ubldeft\ can now be made explicit by introducing
\eqn\subS{
 S^F_{ab}=\pm\prod_{\b\in\G_b^+}\usbl{u(\phi_a,\beta)+1}^{(\l_a,\b)}.}
With this definition,
\eqn\bt{
 S^{\min}_{ab}=\CR_1S^F_{ab} \quad,\quad
 S^B_{ab}={\CR_1S^F_{ab}\over\CR_{1-B}S^F_{ab}}~.}
Note, the sign in \subS\ is left undetermined by the requirement that
\bt\ should hold; indeed $S^F_{ab}$ can be thought of as a skewed
`square root' of $S^{\min}_{ab}$. As long as the function is used merely
as a
book-keeping device, this ambiguity will not cause any problems.
However, it is tempting to speculate about a physical interpretation,
in which case the sign might perhaps be
important and a further subtlety should be considered:
namely, whatever the sign choices made in \subS, the functions
$S^F_{ab}$ will in general fail, by a sign, to satisfy some of the
bootstrap equations holding for $S^{\min}_{ab}$ and $S^B_{ab}$\foot{this
sign was missed in \rRb.}.
The origin of these signs can be seen by re-examining the
proof, given in \rDf, that the bootstrap equations hold for
the functions defined in \myS{}. Since the bootstrap equations often
involve shifts in arguments more general than the symmetrical operation
defined in \Rdef, the first step must be to write \myS{}\ using
functions which transform into themselves under the {\it single} shift
$\CT_y$, defined by
\eqn\Tdef{(\CT_yf)(\t)=f(\t+i\pi y/h).}
This property does not hold for $\usbl{x}$ defined in \usbldef, but
decomposing it as $\usbl{x}=\usbl{x}_+/\usbl{{-}x}_+$, where
\eqn\sbldef{\usbl{x}_+(\t)=
 \sinh\bigl({\theta\over 2}+{i\pi x\over 2h}\bigr)~,}
results in a building block which satisfies
\eqn\Tprop{\CT_y\usbl{x}_+=\usbl{x+y}_+~.}
Similarly $\ubl{x}_+$ can be defined by replacing $\usbl{.}$ by
$\usbl{.}_+$ in either \ubldef\ or \mubldef, in terms of which \myS{}\
becomes
\eqn\myoldS{
 S_{ab}=\prod_{\b\in\G_b}\ubl{u(\phi_a,\beta)+1}_+^{(\l_a,\b)}.}
(The root $\b$ now runs round the whole orbit $\G_b$, rather than just
the positive half as in \myS{}; to see the equivalence of \myoldS\ and
\myS{}\ needs a few root system identities, which can be found in \rDf.)
Now an intermediate step in the proof of the bootstrap equations
involves re-expressing the product appearing in \myoldS, and for this,
the relation $\ubl{x{+}2h}_+=\ubl{x}_+$ must be used. It is this
property which fails by a sign when the same trick is attempted for
$S^F_{ij}$, since $\usbl{x{+}2h}_+=-\usbl{x}_+$. Thus if the manipulations
involve shifting an odd number of blocks $\usbl{x}_+$, an extra minus
sign will appear, and even the freedom of signs allowed by \subS\ is
not enough to eliminate all these signs from the bootstrap equations
(this can already be seen in the $A_2$ case).
There may be some interesting features hidden in these signs,
but the question will not be
pursued here. The signs in any event disappear once the logarithmic
derivative has been taken, and so do not affect the TBA equations.

To discuss these logarithmic derivatives, set
$\p_f(\t)=-i{d\over d\t}\log f(\t)$.
Following from this definition,
\eqn\pprop{\p_{fg}=\p_f+\p_g,\quad\p_{f^{-1}}=-\p_f,
 \quad\p_{\CR_yf}=(\CT_y+\CT_{-y})\p_f.}
With the shorthands $\p^F_{ab}$, $\p^{\min}_{ab}$ and $\p^B_{ab}$
defined in the obvious way, combining \bt\ with \pprop\ yields
\eqn\pS{\p^B_{ab}=\p^{\min}_{ab}-\CT_{1-B}\p^F_{ab}-\CT_{B-1}\p^F_{ab}.}
For consistency with \rRb, we write $\p^F_{ab}$ as $\psi_{ab}$.
Then, putting
\eqn\tdef{\t_0={i\pi\over h}(1-B),}
equation \pS\ can be rewritten as
\eqn\pres{\p^B_{ab}(\t)=\p^{\min}_{ab}(\t)-\psi_{ab}(\t+\t_0)
 -\psi_{ab}(\t-\t_0).}
This relation will be useful in the treatment of the TBA equations.

\newsec{The TBA equations}

The TBA equations for the affine Toda models have the standard form
for a purely elastic scattering theory\ts\rZg. Writing the mass
ratios as $\hm_a{=}m_a/m_1$, a system living on a circle of finite length
$R$ depends on the scale through the quantity ${1\over
2}m_1R\equiv e^x$, and the ultraviolet limit corresponds to sending
$x\rightarrow -\infty$. To avoid the complications caused by the fact that
$x$ will almost always be negative, this parameter
will be swapped for $y=-x$.
The TBA equations for the pseudoenergies $\ep_a(\t)$, encoding all of
the finite-size effects, are
\eqn\TBA{\ep_a(\t)=2\hm_ae^{-y}\cosh(\t )-{1\over 2\pi}\sum_{b=1}^r
\p^B_{ab}*L_b(\t),}
where $L_a(\t)=\log(1+e^{-\ep_a(\t)})$ and $*$ denotes the rapidity
convolution:
\eqn\convdef{\p*L(\t)=\int^{\infty}_{-\infty}d\t'\p(\t-\t')L(\t').}
The energy terms mentioned earlier are the functions
$2\hm_ae^{-y}\cosh(\t )$; they can be attached to the nodes of the
Dynkin diagram via the
association\ts\NRF\rFb\Fb\refs{\rKMa,\rBCDSf,\rFb}\ between the
Toda particle masses and the Perron-Frobenius eigenvector of the Cartan matrix.
The effective central charge, defined to be $-\pi/6R$ times the
finite size ground state energy $E(R)$, is given by
\eqn\cdef{c(y,\t_0)={6\over\pi^2}\sum_{a=1}^r
 \int^{\infty}_{-\infty}d\t\hm_ae^{-y}\cosh(\t )L_a(\t),}
the $\t_0$-dependence of this function entering via the $B$-dependence
of the kernel function $\p^B_{ab}$ in \TBA.

To gain a general understanding of the solutions to \TBA, the
form of this kernel should be discussed. Since
\eqn\psbldef{\p_{\usbl{x}}(\t)=
{1\over 2}\coth\({\t\over 2}+{i\pi x\over 2h}\)-
{1\over 2}\coth\({\t\over 2}-{i\pi x\over 2h}\)~,}
it follows from the definitions \myS b\ and \subS\ (and the reality of
the numbers $u(\p_a,\b)$\tt ) that for {\it real} $\t$
the functions $\p^{\min}_{ab}(\t)$ and
$\psi_{ab}(\t)$ are strongly peaked about $\t=0$, and decay like
$e^{-|{\t}|}$ outside an interval of size $O(1)$ centred at the
origin. Thus so long as $L(\t)$ is reasonably well-behaved on the real
$\t$-axis,
the values of $\p^{\min}*L(\t)$ and $\psi *L(\t)$ are governed by the
values taken by $L(\t')$ with $\t'$ close to $\t$, and the convolution
\convdef\ has a `local' character. (To gain an intuition for this, it
is helpful to think of replacing the functions $\phi^{\min}$ and
$\psi$ by two delta-functions, each normalised to preserve the
corresponding overall integral;
indeed we shall be working in asymptotic regions where this seems to
be a valid approximation to make.)
For real values of the coupling constant $\b$, the shift $\t_0$ in
\pres\ is purely imaginary, and the above line of reasoning goes
through for $\p^B$ as well, with the conclusion that the full TBA equation
also has this local character.
But to incorporate Zamolodchikov's roaming idea, and in particular to
introduce another scale into the problem, it is necessary to be a little
more general.
This should involve relaxing the requirement that $\b$ be real, but on
general grounds it might be expected that the S-matrix elements should
continue to be real analytic -- that is,
$S_{ab}(\t)$ should still be real for $\t$ on the imaginary axis.
{}From the definition \ubldef\ of the basic block, {\it any} real value
for $B$ satisfies this condition, corresponding to $\b$ real if $B$
is between $0$ and $2$, and $\b$ purely imaginary otherwise. A little
less obvious, but clear enough once \ubldeft\ is considered, is that
values of $B$ such that $(1{-}B)^*=-(1{-}B)$ are also permitted (here,
$^*$ denotes complex conjugation). That is, $1{-}B$ should be imaginary,
and $\t_0$ real. In turn, this requires that $\b/(2\surd\pi)$ should
lie on the unit circle in the complex plane. An appealing
feature of the S-matrix elements for such values of $\b$ is that
(unlike the situation for many imaginary values of $\b$) no extra poles are
introduced into the physical strip, and so the interpretations of all
the physical poles, bound states and bootstrap equations remain
unchanged. It also connects with an interesting physical
interpretation of the roaming behaviour, a speculative point that
will be described in the concluding section. For now, the important point is
that this analytic continuation
changes the `locality structure' of the TBA equations
dramatically, as can be seen by using \pres\ to rewrite \TBA:
\eqnn\rTBA
$$\eqalignno{\ep_a(\t)={}&2\hm_ae^{-y}\cosh(\t )&\cr
 &\quad -{1\over 2\pi}\sum_{b=1}^r\[\p^{\min}_{ab}*L_b(\t)
  -\psi_{ab}*\(L_b(\t{-}\t_0)+L_b(\t{+}\t_0)\)\].&\rTBA\cr}$$
The $\p^{\min}$ term continues to `see' $L_b$ in the region
$\t'\approx\t$, but for the $\psi$ term it is now the values taken by
$L_b$ at $\t\pm\t_0$, rather than at $\t$ itself, that are important.
Furthermore, the equation now contains two scales,
namely $y$ and $\t_0$. It is the interaction of these two scales which
produces the staircase behaviour, the controlling parameter
being their ratio, $y/\t_0$. If this ratio is kept fixed while $y$ and
$\t_0$ tend to infinity, then generally the effective central charge
$c(y,\t_0)$
will tend to the central charge of one of the $W_g$-minimal models --
precisely which one depends on the chosen value of $y/\t_0$.
Different asymptotic behaviours are separated by
integer or half-integer values for this ratio. If on the other hand
$y$ is increased while $\t_0$ is kept
fixed, to study the behaviour towards the ultraviolet
of one particular theory at
one particular value of the coupling constant, then the changing value
of $y/\t_0$  results in a series of different values for the
effective central charge, with a cross-over at each integer and
half-integer. The rest of this section is devoted to some analytical
arguments in support of these claims, leading to the expectation
that the set of solutions to \rTBA\ at given (sufficiently large)
$\t_0$ do indeed form a staircase, with the central charge of a
\coset k {k+1}\ coset conformal field theory at the $k\th$ step.

As a first step, it seems reasonable to suppose that for fixed $y$ the
function $L_b(\t)$ is bounded.  As soon as $|\t|$ is large enough for
the first (energy) term in \rTBA\ to
contribute, the exponential growth of the hyperbolic cosine will
quickly swamp any influence of the convolution terms. Thus, for
$\t\gg y$ or $\t\ll -y$, $\ep_a(\t)$ is closely approximated by the
functions $\ep^+_a(\t)$ or $\ep^-_a(\t)$ respectively, where
\eqn\eppm{\ep^{\pm}_a(\t)=\hm_ae^{-y\pm\t}.}
Correspondingly, $L_a(\t)$ suffers a double-exponential decay beyond
$|\t|>y$ and is soon approximately zero.
This information is already enough to see why the $k{=}1$
fixed point is expected to control the effective central charge for
$y\ll\t_0/2$. The reason is simply that for such a value of $y$, the
$\psi_{ab}$ term never contributes significantly to \rTBA\ --- for
$|\t|\gg y$, the energy term dominates and {\it all} the
convolution terms can be ignored, while for $|\t|\leq y$, both $|\t-\t_0|$
and $|\t+\t_0|$ are larger than $y$, and hence $L_b(\t{-}\t_0)$ and
$L_b(\t{+}\t_0)$ are essentially zero, and the $\psi_{ab}$ term in
\rTBA\ can again be dropped.
The resulting `reduced' set of equations is just that
for $S^{\min}_{ab}(\t)$, shown in refs.\ts\refs{\rZg{--}\rKMa}\
to have the central charge of the \coset 1 2\ conformal
field theory as the limiting ($y\rightarrow\infty$) value of $c(y)$.
Of course, $y$ must be less than $\t_0$ for the above
treatment to be valid, but if $\t_0$ is taken large enough then
the relevant central charge is approached well before the approximations
break down.

At this stage it would be possible to continue by discussing
what happens as $y$ increases beyond $\t_0/2$, the values of the
function $L_a(\t)$ near $y$ starting to influence those at $-y$,
and so on. Such a treatment would mimic that already given by
Al.\ts B.\ts Zamolodchikov in \rZo. However, at least if the
aim is merely to uncover the sequence of central charges visited by
the trajectory, it seems better to think of the above discussion just as
a warm-up exercise, and now to proceed directly to the general situation.

An assumption is needed, generalising already-observed behaviour
of `local' TBA equations, on the form of $L'_a(\t)$. For
$\t\ll -y$, $L$ and hence also $L'$ are approximately zero, due to the
dominance of the energy term in \rTBA. Then after a transition around
$\t=-y$ (a kink) during which $L'\neq 0$, $L$ should settle down to
some (not necessarily zero) constant value, and $L'$ return to zero.
In the case of a local TBA equation, this is all that happens until
$\t=+y$ is reached, at which point there is another kink and $L$
returns to zero. However, in the case of \rTBA, the non-local
terms in the
convolution serve to link the values of $L$ and hence $L'$
separated at a distance $\t_0$, and the `seed' kink at $\t=-y$
would be expected to propagate in and cause further kinks at $-y{+}\t_0$,
$-y{+}2\t_0$, and so on. Similarly, the kink at $\t=+y$ results in other
kinks at $\t=y{-}\t_0$, $y{-}2\t_0$, propagating in from the right.
This process does not form any kinks for $|\t|>y$
since in any event the convolutions are swamped by the energy term in
this region.

We can try to make this a little more precise by taking the derivative of
\rTBA\ with respect to $\t$, and substituting $\ep'=-(1{+}e^{\ep})L'$.
This gives an equation which couples the values of $L'$ at $\t$,
$\t{+}\t_0$ and $\t{-}\t_0$. Repeating this at $\t{+}\t_0$,
$\t{+}2\t_0$, and so on couples $L'(\t)$ with $L'(\t+n\t_0)$ for any
$n\in\Z$; furthermore so long as we remain in the region
$|\t{+}n\t_0|<y$, the inhomogeneous terms
($2\hm_ae^{-y}\sinh(\t{+}n\t_0)$\ts ) can be dropped. In addition, if
$|\t{+}n\t_0|>y$, then $L'(\t{+}n\t_0)=0$ is forced (up to exponentially small
corrections) by \eppm. Thus for $\t\not\approx\pm(y{-}n\t_0)$ ($n\in\Z$),
we expect $L'(\t)\sim 0$.  The picture is more
complicated at the `transition regions' around $\t=\pm y$, where the
energy term and the convolutions are in equal competition. The
resulting non-zero values for $L'(\t)$ at $\pm y$ then propagate in
to $-y+\t_0$, $y-\t_0$ and so forth. However, for this
to be any more than a plausibilty argument, it would have to be shown
that the set of equations coupling together the values of $L'$ at
intervals of $\t_0$ were non-singular in a suitable sense, and also
that the $\t$-dependent factors of $-(1+e^{\ep})$ did not spoil this.
Nevertheless, it provides some support for the claim that $L'(\t)$ is close
to zero for $\t$ not equal to $\pm(y{-}n\t_0)$, with $n=0,1...$ and
$y-n\t_0>-y$, and that $L(\t)$ is thus a series of plateaux,
separated by kinks at these points. (Note, the kink regions should each
have a size of order one, and so taking $y$ large while preserving
the value of $y/\t_0$ enables their size relative to $y$ and $\t_0$ to be
made as small as desired.) In the generic situation, kinks propagating in
from the left will miss those propagating in from the right, the two
sets of kinks being interlaced along the $\t$-axis.
This only fails to be true if $y/\t_0$ is near an integer or half-integer;
this should correspond to a crossover in the critical behaviour.
Otherwise, take $y$ to lie between
$(k{-}1)\t_0/2$ and $k\t_0/2$ for some integer $k$.
Working in from $\t=-\infty$, the first kink to be encountered is the
seed kink at $\t=-y$; then a descendant of the rightmost kink
is found at $\t=y-(k{-}1)\t_0$; then a descendant of the leftmost
kink at $-y+\t_0$; and so on.
Between these $2k$ kinks, there are $2k{-}1$ plateaux, within each of
which
the $L_a(\t)$, and hence also the $\ep_a(\t)$, are approximately constant.
The $i\th$ plateau, lying between the $i\th$
and $(i{+}1)\th$ kinks, is centred on $z_i=(i{-}k)\t_0/2$, and
the approximately constant value of $\ep_a(\t)$ along this
plateau will be denoted $\ep^i_a\equiv\ep_a(z_i)$.
These constants are easily found, since
constant terms can be pulled out of the convolutions in \rTBA, leaving
just the overall integrals of $\p^{\min}$ and $\psi$ as
proportionality factors. From \rKMa\ and \rRb, these are
given by
\eqn\Nvals{
{1\over 2\pi}
\int^{\infty}_{-\infty}d\t\p_{ab}^{\min}(\t)=\delta_{ab}-2C^{-1}_{ab}
\quad ;\quad
{1\over 2\pi}
\int^{\infty}_{-\infty}d\t\psi_{ab}(\t)=-C^{-1}_{ab},}
where $C_{ab}$ is the Cartan matrix of the non-affine Dynkin diagram for
$g$. (These formulae can also
be proved directly from \myS b\ and \subS\ts\rDf.)
Considering equation \rTBA\ for $\t=z_i$ results in the following
consistency condition for the numbers $\ep^i_a\,$:
\eqn\epcond{\ep_a^i=\sum_{b=1}^r \[\( 2C^{-1}_{ab}-\delta_{ab}\)\log
(1{+}e^{-\ep_b^i})-C^{-1}_{ab}\(\log(1{+}e^{-\ep^{i-2}_b})+
\log(1{+}e^{-\ep^{i+2}_b})\)\].}
If $z_i$ lies towards an end of the chain, that is if
$i=1,2,2k{-}2$ or $2k{-}1$, the
pseudo-energy there interacts with $\ep^{\pm}(\t)$, already fixed by
\eppm; to the approximation to which we are working here (and
increasing $y$ while keeping $y/\t_0$ fixed if necessary to sharpen things
up), these functions can be taken
to be infinite at the points at which they need to be evaluated. Thus,
with the
boundary conditions that $\ep^{-1}=\ep^{0}=\ep^{2k}=\ep^{2k+1}=\infty$,
\epcond\ should hold for all $i=1\dots 2k{-}1$.
Note that $\ep^i$ interacts with $\ep^{i\pm 2}$ rather
than with its immediate neighbours $\ep^{i\pm 1}$, a
consequence of \rTBA\ and the fact that $z_i\pm\t_0=z_{i\pm 2}$, not
$z_{i\pm 1}$. Thus the constants $\{\ep^i_a\}$ divide into
two sets, $\{\ep_a^{2j}\}$ and $\{\ep_a^{2j-1}\}$, values in each set only
interacting with others in the same set or, at the ends, with the
(effectively infinite) boundary values determined by \eppm. This can be used
to put \epcond\ into a more familiar form: setting
$x^j_a=\exp(-\ep^{2j-1}_a)$ and $y^j_a=\exp(-\ep^{2j-2}_a)$, the
consistency conditions become
\eqn\xycond{\eqalign{
x^j_a=\prod^r_{b=1}&\(1+x^j_b\)^{\delta_{ab}-2C^{-1}_{ab}}
\(1+x^{j-1}_b\)^{C^{-1}_{ab}}%
\(1+x^{j+1}_b\)^{C^{-1}_{ab}},~~j{=}1...k\cr
y^j_a=\prod^r_{b=1}&\(1+y^j_b\)^{\delta_{ab}-2C^{-1}_{ab}}
\(1+y^{j-1}_b\)^{C^{-1}_{ab}}%
\(1+y^{j+1}_b\)^{C^{-1}_{ab}},~~j{=}2...k\cr}}
with the understanding that $x^0_a=x^{k{+}1}_a=y^1_a=y^{k{+}1}_a=0$.
In the $A_1$ case, there is just one particle type and
$C^{-1}_{11}=1/2$; the system reduces to that given in \rZk,
with $x^i_1$ and $y^i_1$ becoming
$x_i$ and $y_i$ in Zamolodchikov's notations.
On the other hand, \xycond\ is
also the specialisation to $l=1$ of the constraints for the
\ccoset k l {k+l}\ TBA system proposed in \rRb: to see this,
$x^i_a$, $y^i_a$ above should be replaced by $Y^a_i(-\infty)$,
$Y^a_{i-1}(+\infty)$ respectively.
Of course, in the two above-mentioned papers the `morality' is rather
different from here,
in that (for fixed particle type $a$) $x_a^j$ and $y_a^j$ are
the asymptotic values at $\pm\infty$ of a number of {\it different}
functions, indexed by $j$; here, they are the intermediate values of a
{\it single} function, $\exp(-\ep_a(\t))$.
This changes the spirit of the central charge calculation, although
the underlying structure is the same.

One more piece of notation is needed. The interleaving of
plateaux and kinks described above can be summarised in the following
inequalities:
\eqn\ineqdef{
\eqalign{z_0<-y<z_1<y{-}(k{-}1)\t_0<&z_2<-y{+}\t_0<z_3<\dots\cr
 &\dots<z_{2k-1}<-y{+}(k{-}1)\t_0<z_{2k}\cr}}
(recall, $z_i=(i{-}k)\t_0/2$, and $(k{-}1)\t_0/2<y<k\t_0/2\,$). Define
$K_i$ to be the interval $[z_{i-1},z_i]$, so that the $i\th$ kink of
$L_a(\t)$ lies in $K_i$. (For $i\neq 1, 2k$ this is also a kink for
$\ep_a$, while for $i=1$ or $2k$ it marks the transition of $\ep_a$ to
the exponentially growing behaviour described by \eppm.) For $y/\t_0$
sufficiently far from the crossover (integer or half-integer) values,
the kinks occur well inside the regions $K_i$, and by the time the
boundaries of these regions are reached, the functions $L_a$ have
settled down to constants. Were it not
for the non-local nature of \rTBA, these kinks would be described by a
set of independent TBA equations, with boundary conditions,
effectively at $\pm\infty$, determined by \xycond. However, the nonlocality
couples the equation for the kink in $K_i$ to those for the kinks in
$K_{i\pm 2}\,$; to leading order these coupled equations are exactly
those proposed in \rRb. In this way, the `tower' structure of the TBA
systems in that paper can be traced back to the block structure of the
affine Toda S-matrices, the key relation being \pS.

This would be enough to verify our claims; however for completeness
the remainder of section concludes the computation of the
function $c(y,\t_0)$ directly in the context of the staircase model, continuing
to work in the asymptotic region where both $y$ and $\t_0$ are large,
with $(k{-}1)/2<y/\t_0<k/2$.

Equations \cdef\ and \rTBA\ must be re-examined with an eye to
their asymptotic behaviour. The integration range $[-\infty,\infty]$
in the expression \cdef\ for the effective central charge $c(y,\t_0)$
can be restricted to $[z_1,z_{2k}]=\cup_{i=1}^{2k}K_i$, since
the double-exponential decay of $L_a(\t)$ causes the integrand to be
vanishingly small outside this region. Within this region,
\eqn\coshasymp{ 2\hm_ae^{-y}\cosh(\t)\sim
\cases{\hm_ae^{-y-\t},&$\t\in K_1\, ;$\cr
                    0,&$\t\in K_2\dots K_{2k-1}\, ;$\cr
       \hm_ae^{-y+\t},&$\t\in K_{2k}\, .$\cr }}
Thus the only direct contribution to $c(y,\t_0)$ is from the kinks in $K_1$
and $K_{2k}\,$.  Writing $c(y,\t_0)=c_-+c_+$, where
\eqn\cpmdef{c_{\pm}={3\over\pi^2}\sum_{a=1}^r
 \int^{z_{2k}}_{z_1}d\t\hm_ae^{-y\pm\t}L_a(\t)\, ,}
and noting by symmetry that $c_+=c_-$, all that remains is to
calculate
\eqn\cmdef{c_-\sim {3\over\pi^2}\sum_{a=1}^r
 \int_{K_1}d\t\hm_ae^{-y-\t}L_a(\t).}
This is essentially standard, the only modifications needed to the
cases in \refs{\rZk,\rMk,\rRb} being, crudely speaking, that integration
by parts should be replaced by considerations based on the
`locality' of $\phi^{\min}(\t)$ and $\psi(\t)$.
The relevant property is the following: let $f(\t)$ be any even
function of $\t$ with its support concentrated at the origin
(by this we mean that the overall integral of $f(\t)$ receives a
significant contribution only in a region near $\t{=}0$, the size of
which is asymptotically small in comparison with $\t_0$ and $y$;
both $\phi^{\min}(\t)$ and $\psi(\t)$ satisfy this condition).
If $A(\t)$ and $B(\t)$ vary slowly (on the scale of the support
of $f$) at the ends of the interval $K_i=[z_{i-1},z_i]$, then
\eqn\partprop{\int_{K_i}d\t f'{*}A(\t)B(\t)
 =-\int_{K_i}d\t f'{*}B(\t)A(\t)+\(\intt d\t
f(\t)\)\,\bigl[A(\t)B(\t)\bigr]^{z_i}_{z_{i-1}}\,.}
If $f(\t)=\delta(\t)$, this formula really is just integration by
parts; a proof of its more general validity is relegated to appendix
A.

Returning now to the calculation of the effective central charge,
the necessary property of the solutions of the TBA equations, allowing
for the exact evaluation of \cpmdef\ in the asymptotic region of
interest, is found by
taking the derivative of \rTBA, multiplying by $L_a(\t)$, integrating
$\int_{K_i}d\t$ and finally summing on $a$. In an abbreviated notation,
this yields
\eqn\dTBAprop{\CJ^i= {\pi^2\over 3}\CC^i-\CI_0^i+\CI_-^i+\CI_+^i\,}
where
\eqna\CIdefs
$$\eqalignno{
\CJ^i&=\sum_{a=1}^r\int_{K_i}d\t\ep'_a(\t)L_a(\t)=
\sum_{a=1}^r\int_{\ep_a^{i-1}}^{\ep_a^i}d\ep\log\(1+e^{-\ep}\)\, ;
&\CIdefs a\cr
\CC^i&= {3\over\pi^2}
 \sum_{a=1}^r\int_{K_i}d\t 2\hm_ae^{-y}\sinh(\t)L_a(\t)\, ;&\CIdefs b\cr
\CI_0^i&={1\over 2\pi}\sum_{a,b=1}^r\int_{K_i}d\t\phi^{\min\,\prime}_{ab}%
 {*}L_b(\t)L_a(\t)\, ;&\CIdefs c\cr
\CI_{\pm}^i&={1\over 2\pi}\sum_{a,b=1}^r\int_{K_i}d\t\psi'_{ab}%
 {*}L_b(\t\pm\t_0)L_a(\t)\, .&\CIdefs d\cr }$$
These terms can be simplified in various ways. First, using the relations
\eqn\sinhasymp{ 2\hm_ae^{-y}\sinh(\t)\sim
\cases{-\hm_ae^{-y-\t},&$\t\in K_1\, ;$\cr
                    0,&$\t\in K_2\dots K_{2k-1}\, ;$\cr
       \hm_ae^{-y+\t},&$\t\in K_{2k}\, ,$\cr }}
and referring back to \cpmdef, we have
\eqn\CCasymp{\CC^i\sim -\delta_{1,i}c_-+\delta_{2k,i}c_+\,.}
(Note how the relations \coshasymp\ and \sinhasymp, by forcing the
effective energy terms to lie at the end of the $A$-type Dynkin
diagram, automatically select the
$l{=}1$ cases among the TBA systems proposed in \rRb.)

Of the remaining pieces, $\CI_0^i$ is just a surface term: applying
\partprop\ to \CIdefs c, swapping the sums on $a$ and $b$ and
recalling that $\phi^{\min\,\prime}_{ab}{=}\phi^{\min\,\prime}_{ba}$
yields
\eqn\iores{2\CI_0^i=\sum_{a,b=1}^r \(\delta_{ab}-2C^{-1}_{ab}\)
\bigl[L_a(\t)L_b(\t)\bigr]_{z_{i-1}}^{z_i}\,,}
where \Nvals\ has also been used for the overall integral
of $\phi^{\min}_{ab}$. A similar manoeuver for \CIdefs d\ gives, for
$1\le i\le 2k$,
\eqn\ipmres{\CI_{\pm}^i=-\CI_{\mp}^{i\pm 2}-\sum_{a,b=1}^rC^{-1}_{ab}
 \bigl[L_a(\t)L_b(\t\pm\t_0)\bigr]_{z_{i-1}}^{z_i}\,,}
while $\CI_{\pm}^i\sim 0$ if $i<1$ or $i>2k$, owing to the
double-exponential decay of $L_a(\t)$. The surface piece makes this
equation a little hard to use; to get rid of it, define
\eqn\itpmdef{\tCI_{\pm}^i=\CI_{\pm}^i+{1\over 2}\sum_{a,b=1}^r
C^{-1}_{ab}\bigl[L_a(\t)L_b(\t\pm\t_0)\bigr]_{z_{i-1}}^{z_i}\,.}
Recalling that $z_i{+}\t_0=z_{i+2}$, \ipmres\ is now
\eqn\itpmres{\tCI_{\pm}^i=-\tCI_{\mp}^{i\pm 2}\,,}
and, as before, $\tCI_{\pm}^i\sim 0$ if $i<1$ or $i>2k$.
Equation \dTBAprop\ becomes
\eqn\dTBApp{\eqalign{\CJ^i={\pi^2\over 3}\CC^i+\tCI_-^i+\tCI_+^i
+{1\over 2}\sum_{a,b=1}^r\Bigl[ L_a(\t)\,\Bigl\lbrace
\bigl(&2C^{-1}_{ab}-\delta_{ab}\bigr)L_b(\t)\cr\noalign{\vskip -8pt}
&{}-C^{-1}_{ab}\bigl( L_b(\t{-}\t_0)+L_b(\t{+}\t_0)\bigr)
\Bigr\rbrace\Bigr]_{z_{i-1}}^{z_i}\cr\noalign{\vskip 4pt}}}
equation \iores\ having been used to substitute for $\CI^i_0$. Now the
term in curly brackets is familiar from \epcond, since it is only ever
evaluated when $\t$ is equal to one of the $z_i\,$:
once the sum on $b$ has been performed, it is just $\ep_a(\t)$.
So, if we set
\eqn\tcjdef{\tCJ^i=\CJ^i-
{1\over 2}\sum_{a=1}^r\bigl[L_a(\t)\ep_a(\t)\bigr]_{z_{i-1}}^{z_i}}
and use \CCasymp\ as well, the result is
\eqn\dTBAppp
{\tCJ^i={\pi^2\over 3}\(-\delta_{1,i}c_-+\delta_{2k,i}c_+\)
+\tCI_-^i+\tCI_+^i\,.}
At $i=1$, this is
\eqn\cmsi{c_-=-{3\over\pi^2}(\tCJ^1-\tCI_-^1-\tCI_+^1)\,.}
Now $\tCI_-^1=-\tCI_+^{-1}=0$, while $\tCI_+^1=-\tCI_-^3$. In
addition, for $1<i<2k$, \dTBAppp\ can be rewritten, via \itpmres, as
$\tCI_-^i=\tCJ^i+\tCI_-^{i+2}\,$. Thus \cmsi\ falls out into a cascade
of terms, one for every other kink starting with the left-most:
\eqn\cmsii{\eqalign{c_-&=-{3\over\pi^2}(\tCJ^1+\tCI_-^3)\cr
&=-{3\over\pi^2}(\tCJ^1+\tCJ^3+\tCI_-^5)\cr
&=\dots\cr
&=-{3\over\pi^2}\sum_{j=1}^k\tCJ^{2j-1}\,,\cr}}
the process terminating when the term $\tCJ^{2k+1}=0$ is encountered.
Note that the last line only depends on $y$ and $\t_0$ via
the value of $k$ that their ratio entails.
To calculate $\tCJ^{2j-1}$, we write it as
$\sum_{a=1}^r\tCJ^{2j-1}_a$, where
\eqn\tcjadef{\tCJ^{2j-1}_a=\CJ^{2j-1}_a-
{1\over 2}\bigl[L_a(\t)\ep_a(\t)\bigr]_{z_{2j-2}}^{z_{2j-1}}}
and
\eqn\cjadef{\eqalign{\CJ^{2j-1}_a
&=\int_{\ep_a^{2j-2}}^{\ep_a^{2j-1}}d\ep\log\(1+e^{-\ep}\)\cr
&=\dilog\(-e^{-\ep_a^{2j-1}}\)-\dilog\(-e^{-\ep_a^{2j-2}}\)\,.\cr}}
Here, $\dilog(z)=\int_z^0dt\log(1-t)/t$ is the `usual'
dilogarithm function (for more information, see ref.\ts\RF\rLc\Lc\tt).
The surface part in \tcjadef\ turns this into
the Rogers dilogarithm, $\rdilog(z)=-{1\over
2}\int_0^zdt\({\log(1-t)\over t}+{\log t\over 1-t}\)\,$, via the
relation
\eqn\rdlident{\rdilog\({x\over 1+x}\)=-\dilog(-x)-\log(1{+}x)\log x\,,}
valid for $x>-1$\ts\rLc. With the earlier definitions
$x^j_a=e^{-\ep^{2j-1}_a}$, $y^j_a=e^{-\ep^{2j-2}_a}$,
equation \tcjadef\ becomes
\eqn\tcjares{\tCJ^{2j-1}_a=-\rdilog\({x^j_a\over 1+x^j_a}\)+
\rdilog\({y^j_a\over 1+y^j_a}\)\,.}

The effective central charge is $c_-{+}c_+=2c_-$. Subject to the hypotheses
on the plateau-like behaviour of $L_a(\t)$, we have now shown that
\eqn\resulti{\lim_{{y,\t_0\to\infty\atop (k-1)<2y/\t_0<k}}c(y,\t_0)=
c_k={6\over\pi^2}\sum_{j=1}^k\sum_{a=1}^r
\[\rdilog\({x^j_a\over 1+x^j_a}\)-\rdilog\({y^j_a\over 1+y^j_a}\)\]\,,}
where $x^j_a$ and $y^j_a$ are the real solutions of \xycond\ (or zero in
the case of $y^1_a\,$, corresponding to the first kink). In taking
this limit, the ratio $y/\t_0$ is assumed to be kept fixed. If instead
$\t_0$ is fixed at some (suitably large) value and $y$ is increased,
$c(y,\t_0)$ will run through the values $c_k$ in turn. The plateau
structure should break down, and a cross-over occur, whenever the
interval between a pair of kinks is of the same order as the kink
size. The intervals
between the kinks are of lengths $2y-(k{-}1)\t_0$ and $k\t_0-2y$,
while the kinks themselves have an extent of order one. Thus the
cross-overs should occur while $2y$ is
in a region of order one about each integer multiple of $\t_0$.
(Examining the graphs in ref.\ts\rZo, the correct size would seem to
be about five, a value completely consistent with the form \psbldef\ of
the functions $\p_{(x)}\,$. The important point is that the sizes of
these cross-over regions depend neither on $\t_0$ nor on $y$.)

To find numerical values for the $c_k$, we use the following sum
rule\ts\RF\rBRa{\Kg\semi\BRa},
which holds whenever the constants $x^i_a$ satisfy a `chain' of
relations of the form \xycond, of length $k$:
\eqn\sumrule{ \sum_{j=1}^k\sum_{a=1}^r
\rdilog\({x^j_a\over 1+x^j_a}\)={\pi^2\over 6}{rhk\over h+k+1}\,.}
(So for the constants $y^i_a$ in \xycond, $k$ should be replaced by
$k{-}1$ here.) For the simply-laced Lie algebras, the Coxeter number $h$
is equal to the dual Coxeter number $\tilde h$, so the right-hand side of
\sumrule\ is in fact the central charge for a theory of Gepner
parafermions\ts\RF\rGc\Gc\ at level $k{+}1$. Since we do not know of any
general proof of these relations, we
checked them numerically (to eight-digit final precision)
for all simply-laced Lie algebras of rank $\leq 8$,
for $k=1\dots 8$. With the relation
$$\rdilog\({x\over 1+x}\)={\pi^2\over 6}-\rdilog\({1\over 1+x}\)\,,$$
this gives
$$c_k=r(h+1)\[{1\over h+1}+{k\over h+k}-{k+1\over h+k+1}\]\,,$$
a result which reproduces the central charge of the \coset k {k+1}\
coset model.

\newsec{Conclusions and speculations}

The hopping nature of the solutions to equation \rTBA\ has now been
established in a fair amount of detail, but the physical meaning of
this result is much less clear. While we have no final answers to this
question, we give in this concluding section some speculations
on the Lagrangian form of the staircase models which might be worthy
of further investigation.

The usual Lagrangian density for affine Toda field theory can be
written as
\eqna\todlag
$${\cal L}={1\over 2}(\partial\p)^2-V(\p,\b)\eqno\todlag a$$
where
$$V(\p,\b)={m^2\over\b^2}\sum^r_{i=0}n_ie^{\b\a_i.\p},\eqno\todlag b$$
the sum in the second equation running over all the simple roots
$\a_1\dots\a_r$ of (non-affine) $g$, together with
$\a_0=-\sum^r_1n_i\a_i$, the negative of the highest root. So long as
$\b$ is real, the potential has a unique minimum at $\p=0$, and a
purely elastic scattering theory of $r$ different types of bosonic
particles results\ts\refs{\rAFZa{--}\rDDa}.

However the potential \todlag b\ has also been
discussed at purely imaginary values of
$\b$\ts\NRF\rEYa\EYa\NRF\rHMa\HMa\NRF\rHd\Hd\refs{\rEYa{--}\rHd},
the main aim being to find a
Lagrangian basis for the $\p_{(id,id,adj)}$-perturbed coset models
(generally thinking of the perturbation as being in the massive direction).
In such cases the unitarity of \todlag a\ is doubtful, and
some form of truncation has to be invoked to restrict the space of
states to a positive-definite subspace. The physical motivation for this
procedure is unclear, and so it might be preferable to look for an
alternative continuation of the real coupling-constant Lagrangian for which
the manifest reality of the potential was preserved.

The simplest
way to create a real potential from \todlag b\ is just to add
its complex conjugate, forming $V'(\p,\b)=V(\p,\b)+V(\p,\b^*)$. (We
assume that the field $\p$ remains real here.) However, the
non-linearity of the Toda field equations means that at the classical
level there is no reason to believe the potential $V'$ to be
integrable. But at the quantum level, for suitably well-chosen values
of $\b$, things may be different. The basis of this expectation is the
strong-weak coupling duality of the affine Toda field theories, a
quantum effect seen for the S-matrices in the fact that
(from equation \ubldef\tt) $S^B_{ab}=S^{2-B}_{ab}$. Since
$B(\b)=2\b^2/(\b^2+4\pi)$, the interchange between the two
possibilities is effected by the following `duality' transformation on
the coupling constant:
\eqn\bdual{\b~\rightarrow~ \tilde\b=4\pi/\b\,,}
relating the strong and weak coupling regimes. At the more fundamental
level of light-cone canonical quantization, it has also been
argued\ts\NRF\rDDb\DDb\ref\vinny{V.\ts Fateev, unpublished;
mentioned in ref.\ts\rDDb}\ that the Hamiltonians arising from
$V(\p,\b)$ and $V(\p,\tilde\b)$ actually commute -- a result from which
the equality of the strong- and weak- coupling S-matrices follows
automatically. This motivates the idea that a linear combination of
$V(\p,\b)$ and $V(\p,\tilde\b)$ might be quantum integrable. One
possibility is to form the `self-dual' potential
$V''(\p,\b)=V(\p,\b)+V(\p,\tilde\b)$, and in fact
analogous non-affine potentials have already
been examined by Mansfield\ts\RF\rMn\Mn, though with rather different
motivations. Here, the hope is to construct a real continuation of
\todlag {}\ which
retains integrability at the quantum level, and we now see that this
might be possible if $V'(\p,\b)=V''(\p,\b)$, that is if
$\b^*=\tilde\b$. The allowed values of $\b$ are thus
\eqn\betagamma{\b=2\sqrt{\pi}e^{i\g},}
and the `roaming' analytic continuation has been
recovered, from an alternative perspective. As a function of the
(real) variable $\g$, the proposed potential is
\eqn\vsdual{V''(\p,\g)=
{m^2\over 2\pi}\sum^r_{i=0}n_ie^{2\sqrt{\pi}\cos\g\a_i.\p}
\cos\[(2\sqrt{\pi}\sin\g)\a_i.\p+2\g\]\,.}
Taking $\g=0$ recovers the real-coupling Toda theory at its self-dual
point, while (since $\t_0={\pi\over h}\tan\g$) the large-$\t_0$ limit
corresponds to $\g\to\pi/2$. In this limit each term in the sum
\vsdual\ is a product
of a slowly-growing exponential with a rapidly-varying cosine,
and near the origin the total is well approximated by
\eqn\vsper{ -{m^2\over 2\pi}\sum^r_{i=0}n_i
 \cos\[(2\sqrt{\pi}\sin\g)\a_i.\p+2\g\]\,.}
We would like to think of this as the `unrestricted' potential, a
manifestly real
analogue, for the higher-rank algebras, of the sine-Gordon potential.
In the $A_2$ case, for example, the potential \vsper\ has minima lying
at the vertices of a triangular lattice. If these are the
classical vacua, then (at least from a visual examination of the
potential) kinks should exist, joining neighbouring vacua, one for
each edge of the lattice. This motivates the idea that \vsper\ might
provide a potential for a Landau-Ginzburg description of an
unrestricted SOS-type model associated with the algebra $g$.
Further from the origin, the exponential growth of the other terms in
\vsdual\ becomes important. At first sight this is unfortunate, as an
immediate consequence is that the potential \vsdual\ is not bounded
from below, and so it seems that the theory is not well-defined.
However (at least so long as $\t_0$ is large enough) while the minima
get lower at larger values of $|\p|$, so the saddle-points between
these minima get higher. At a given energy-scale, this may serve to
confine the field to a neighbourhood of the origin, the vacua further
away being effectively decoupled. To take a classical analogy, a particle
starting at the origin of the potential \vsdual\ with {\it any} finite
kinetic energy would be confined to a finite domain in the $\p$-plane,
despite the fact that beyond this domain points can be found at which
the potential energy is arbitrarily negative. The
question is thus whether 1+1 dimensions is near enough to the
classical ($d=\infty$) limit. Assuming that this is true, a rather
appealing picture emerges --- at any given energy-scale, only a
certain number of vacua are accessible to the field $\p$, and the SOS
picture provided by \vsper\ is modified to an effective
Landau-Ginzburg potential appropriate for a restricted SOS-type model
(the possibility of such potentials for the case of $a_2$ is
mentioned in \RF\rZc\Zc\tt). As one progresses towards higher
energies in the UV limit, more and more vacua become available and the
restriction is lifted in stages, in a way which could reproduce the
roaming behaviour already observed at the level of TBA equations.

It has to be admitted that there are many problems with this scenario,
even at the naive classical level presented above. In particular, it
seems hard to ensure that the easing of the restriction happens in
the correct manner, so that a full series of coset models (or rather
their hoped-for
Landau-Ginzburg potentials) is visited. In this regard, it may be
relevant that there is no compelling reason for the
choice of $V''$ given above -- more generally, pick any function $f$,
with complex conjugate $\bar f$. Then $f(\b)V(\p,\b)+\bar
f(\tilde\b)V(\p,\tilde\b)$ is a candidate potential. Clearly it would
be more satisfying to have a physical principle to justify one choice
over all the others. It should also be stressed again that the models
described by \vsdual\ are perforce a long way from their classical
domains, and so (just as for the Landau-Ginzburg picture
of the $c<1$ models\ts\RF\rZp\Zp\tt) classical intuitions should be
treated with caution.

Despite these difficulties, the search for a physical understanding of
the staircase models seems to be very worthwhile. At the level
of the TBA, they already unify and clarify a whole class of
systems; it is not unreasonable to hope for similar insights at the
field theory level as well.

\bigskip\noindent{\bf Acknowledgements}\smallskip\nobreak
We would like to thank the Theory Group at Saclay, where this
collaboration was begun, for their hospitality.
PED is grateful to the European Community for a
grant under the EC Science Programme.
\medskip

\appendix{A}{}
The aim is to show that
$$\int_{K_i}d\t f'{*}A(\t)B(\t)
 =-\int_{K_i}d\t f'{*}B(\t)A(\t)+\(\intt d\t
f(\t)\)\,\bigl[A(\t)B(\t)\bigr]^{z_i}_{z_{i-1}}\,,\eqno\partprop$$
subject to the conditions already mentioned in section 3. First,
expand out the left-hand side as follows:
\eqn\LHSi{\eqalign{\int_{K_i}d\t f'{*}A(\t)B(\t)&=
\int_{K_i}d\t\intt d\t' f'(\t-\t')A(\t')B(\t)\cr
&=\int_{\ri\cup \riii\cup \rv}d\t d\t' f'(\t-\t')A(\t')B(\t),\cr}}
where $\ri$, $\riii$ and $\rv$ are three regions in the $\t,\t'$ plane:
$$\ri=K_i\times [-\infty,z_{i-1}]\,;\quad
  \riii=K_i\times [z_i,\infty]\,;\quad
  \rv=K_i\times K_i\,.$$
Similarly, the first integral on the right-hand side of \partprop\
is, with a trivial change of integration variable,
\eqn\RHSi{\eqalign{-\int_{K_i}d\t' f'{*}B(\t')A(\t')&=
-\int_{K_i}d\t'\intt d\t f'(\t'-\t)B(\t)A(\t')\cr
&=\int_{\rii\cup \riv\cup \rv}d\t d\t' f'(\t-\t')A(\t')B(\t),\cr}}
swapping the order of the integrations (the double integral is absolutely
convergent due to the finite extent of $K_i$) and using the antisymmetry
of $f'$ in the last line. This time the last integral is over the
three regions $\rii$, $\riv$ and $\rv$ where $\rv$ is as before and
$$\rii=[-\infty,z_{i-1}]\times K_i\,;\qquad
 \riv=[z_i,\infty]\times K_i\,.$$
Subtracting \RHSi\ from \LHSi, the second term on the right-hand side
of \partprop\ must be given by
$$\[\int_{\ri}-\int_{\rii}+\int_{\riii}-\int_{\riv}\] d\t d\t'
 f'(\t-\t')A(\t')B(\t).$$
Consider first the integrals over $\ri$ and $\rii$. The locality
of $f'$ means that in these two regions the integrand is only
non-negligible near $(z_{i-1},z_{i-1})$, and the
approximately constant nature of $A(\t')B(\t)$ near this point allows
its replacement in the integrand by $A(z_{i-1})B(z_{i-1})$.
Asymptotically we are also free to replace
$\ri$ by $\ri'=[z_{i-1},\infty]{\times}[-\infty,z_{i-1}]$ and
$\rii$ by $\rii'=[-\infty,z_{i-1}]{\times}[z_{i-1},\infty]$ (it may
help to draw a picture of this), after which the remaining (double)
integral of $f'(\t-\t')$ is easily seen to reduce to $-\intt f(\t)$.
In just the same way the difference between the integrals over $\riii$ and
$\riv$ is well-approximated by $(\intt f(\t)) A(z_i)B(z_i)$, which
completes the demonstration of \partprop.

Note that in the usual TBA, the integrals all run from $-\infty$ to
$\infty$, in which case a naive swap of integration variables
leads to the second term on the right-hand side of \partprop\ being
missed. This would be wrong, the reason being that
the integral over the entire
$\t,\t'$-plane is not absolutely convergent. The correct prescription
is to regulate the $\t$-integral to an interval $K_i$, say, and
then let $K_i\rightarrow [-\infty,\infty]$ at the end, thus recovering
\partprop\ with limits of $\pm\infty$. It is worth stressing that this
result does not depend at all on the `integrable' nature of the TBA
equations, equations which only need to be used {\it once} (in the
form of the relation \dTBAprop\tt) at each step
in the evaluation of the effective central charge.

\listrefs
\bye